# Coherent evolution of signal amplification by reversible exchange in two alternating fields (alt–SABRE)

Andrey N. Pravdivtsev*[a], Nicolas Kempf[b], Markus Plaumann[c], Johannes Bernarding[c], Klaus Scheffler, Jan-Bernd Hövener[a] and Kai Buckenmaier*[b]

| | |
|---|---|
| [a] | Dr. A. N. Pravdivtsev and Prof. Dr. J.-B. Hövener |
| | Section Biomedical Imaging, Molecular Imaging North Competence Center (MOIN CC), Department of Radiology and Neuroradiology, University Medical Center Kiel, Kiel University, |
| | Am Botanischen Garten 14, 24114, Kiel, Germany |
| | E-mail: andrey.pravdivtsev@rad.uni-kiel.de |
| [b] | M. Sc. N. Kempf, Dr. K. Buckenmaier |
| | High-Field Magnetic Resonance Center, Max Planck Institute for Biological Cybernetics, |
| | Max-Planck-Ring 11, 72076, Tübingen, Germany |
| | E-mail: kai.buckenmaier@tuebingen.mpg.de. |
| [c] | Dr. M. Plaumann, Prof. Dr. Dr. J. Bernarding |
| | Institute for Biometrics and Medical Informatics, Otto-von-Guericke University, |
| | Building 02, Leipziger Str. 44, 39120, Magdeburg, Germany |

*Abstract:* **Parahydrogen (pH$_2$) is a convenient and cost efficient source for magnetic resonance signal enhancement. Transient interaction of pH$_2$ with a metal organic complex in a signal amplification by reversible exchange (SABRE) experiment enabled more than 10% polarization for some $^{15}$N molecules. Here, we analyzed a variant of SABRE, consisting of an outer magnetic field alternating between a low field of ~1 µT, where a polarization transfer takes place, and a higher field >50 µT (alt-SABRE). We found effects of both of these fields on amplitude and the frequency of polarization transfer. Deviation of a lower magnetic field from a "perfect" condition of level anti-crossing increases the frequency of polarization transfer that can be exploited for polarization of short-lived transient SABRE complexes i.e. some substrates. Moreover, the coherences responsible for polarization transfer at a lower field persisted during magnetic field variation and continued their spin evolution at higher field with a frequency of 2.5 kHz at 54 µT. The latter should be taken into consideration for an efficient alt-SABRE.**

The amplification of nuclear spin polarization or hyperpolarization enabled many exciting applications in chemistry[1,2], catalysis,[3] biology[4] and, in particular, medical diagnostics.[5,6] In the latter, the major challenge of hyperpolarization is its relatively short lifetime (<1 minute) and a one-time administration of hyperpolarized media (bolus) to the object of interest.[6,7] Although, continuous polarization was demonstrated[8–11] it does not have yet a medical application. Therefore, maximum possible polarization for the given sample size is still desirable and the quest for such methods is ongoing.

Parahydrogen (pH$_2$) based techniques provide one of the most cost-efficient[12] ways of signal enhancement with accessible $^{13}C$ polarization over 50%[13] and $^{15}N$ polarization over 20%.[14] pH$_2$ is a spin-isomer of molecular hydrogen whose nuclear spins are in the singlet spin state $|S\rangle = (|\alpha\beta\rangle - |\beta\alpha\rangle)/\sqrt{2}$ with $\alpha$ and $\beta$ being spin states with parallel and anti-parallel projections of spin-1/2.

Signal amplification by reversible exchange (SABRE)[15] of pH$_2$ with a substrate, S, allows to convert the para-order into observable polarization continuously and within seconds. Although SABRE is a dynamic process and polarization occurs only in the transient Ir-complex, the contact time of pH$_2$ and S, or lifetime of such Ir-complex, $\tau_l$, is sufficiently long to allow polarization transfer by free evolution[15] or RF-pulses.[16] The most efficient polarization transfers under free evolution ("spontaneously") in SABRE happens at a level anti-crossing (LAC) magnetic field $B_{LAC}$[17] where singlet state of pH$_2$ couples with its triplet states and as a result polarization distributes between the crossed states.[18] For the common SABRE systems and pH$_2$ to $^1H$ interaction, $B_{LAC}$ is on the order of 6 mT, while for pH$_2$ to $^{15}N$ interaction, it is on the order of 1 µT[19]. The later experiment was dubbed SABRE in SHield Enables Alignment Transfer to heteronuclei (SABRE-SHEATH).[19]

Recently it was shown that fast and repetitive alternation between $B_{LAC}$[17] and another field (**fig. 1b**), alt-SABRE-SHEATH depicts the J-coupling driven coherent polarization transfer at $B_{LAC}$ and can also provide even higher polarization than SABRE at constant field (**fig. 1a**).[20] Here, we report our observations on tweaking alt-SABRE-SHEATH in a way that the Larmor frequency determines the polarization transfer.

The coherently driven polarization transfer from pH$_2$ to S at $B_{LAC}$ with a frequency on the order of the scalar spin-spin coupling constants (**fig. 2c**) was already experimentally verified by Lindale et al. for relatively long pH$_2$ refreshing times $t_{high}$ > 200 ms at $B_{high}$ = 55 µT.[20] This can be well reproduced using a SABRE model based on either Markov chain Monte Carlo simulations of chemical exchange[20,21] or modified Liouville von Neumann equation with chemical exchange superoperators used here.[21,22]

The active SABRE complex [IrH$_2$S$_2$] (**fig. 2a**) comprises an Ir-complex with ligands, H$_2$ and two equatorial substrates S.[23] The chemical kinetics between S and [IrH$_2$S$_2$] can be described via the exchange rates $k'_a$ and $k_d$.[21,22]

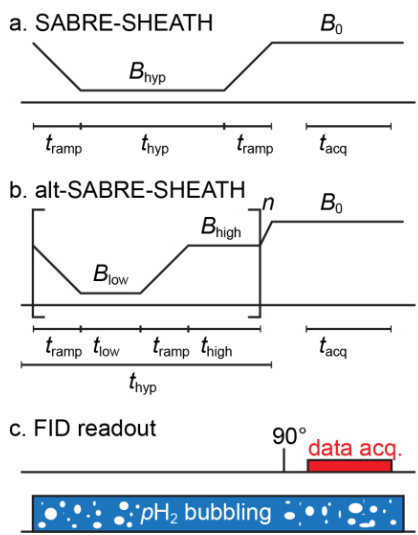

**Figure 1. Sequence schematics.** (a) Magnetic field pattern for SABRE-SHEATHS and (b) for alt-SABRE-SHEATH. (c) A simple free induction decay (FID) readout consisting of a pulse excitation of $^1H$ and $^{15}N$ spins ($B_1$) and signal acquisition at $B_0 \approx 54$ µT with duration $t_{acq}$. Alt-SABRE-SHEATH consists of a hyperpolarization phase where the magnetic field is alternated between $B_{low}$ and $B_{high}$ with duration $t_{low}$ and $t_{high}$ respectively. This part is repeated for $n$ times (total hyperpolarization time $t_{hyp}$). During the entire experiment pH$_2$ is continuously supplied with a flow rate of ≈ 2.5 L/h, which does not harm the spectral resolution at ultra-low fields.

For the sake of simplicity, it can be simplified to an AA'XX' type 4-spin system: AA' are the protons from pH$_2$ and XX' are two $^{15}$N nuclei of two equatorial substrates. The AA' as well as the XX' system can be described using the singlet-triplet basis: $|S\rangle$, $|T_-\rangle = |\beta\beta\rangle$, $|T_0\rangle = (|\alpha\beta\rangle + |\beta\alpha\rangle)/\sqrt{2}$, $|T_+\rangle = |\alpha\alpha\rangle$. Subsequently the spin state of the complete system can be described with a ket vector with two components $|KM\rangle$, $K$ for the AA' and $M$ for the XX' part.

At ≈ 1 µT there is an LAC between two states: $|SS\rangle$ and $|T_-T_+\rangle$ with an energy level splitting on the order of $|\,^2J_{HN}^{trans} - \,^2J_{HN}^{cis}|$, where $^2J_{HN}^{trans}$ describes the J coupling constant for opposite $^1$H and $^{15}$N and $^2J_{HN}^{cis}$ for neighboring.[19] This transition is used in SABRE-SHEATH polarization transfer from pH$_2$ to $^{15}$N. The $|SS\rangle$ state is overpopulated when fresh pH$_2$ binds to [Ir], and at LAC condition the interaction between crossing states $|SS\rangle$ and $|T_-T_+\rangle$ results in a maximum SABRE-SHEATH polarization (**fig. 2b**). In fact, the energy diagram and interaction are more complex even of such a small system (**SI, fig. S9**). Other energy states (and coherences) are also populated and transfer polarization. Important to note is that the total spin projection of the states is constant (e.g. 0 for $|SS\rangle$ and $|T_-T_+\rangle$ states) because scalar spin-spin interaction cannot change the total projection of the state. This means that zero-quantum coherences (or transitions) transfer polarization in this case.

When the magnetic field is higher or lower than $B_{LAC}$ the polarization efficiency is decreasing while the rate of polarization transfer is increasing (or decreasing) since there is no complete match of the Zeeman interaction and J-coupling constants. The alt-SABRE SHEATH reproduces this LAC analysis (**fig. 2c**).

Interestingly, if the magnetic field variation is fast enough ($t_{low}$, $t_{high}$, $t_{ramp} \leq \tau_l$) and $B_{low} \ll B_{high}$, then the zero-quantum coherences are retained and continue to evolve at $B_{high}$ at elevated frequency. By varying $t_{high}$ it was even possible to observe rather faster polarization oscillations with the frequency of $(\gamma_{1H} B_{high} - \gamma_{15N} B_{high})/2\pi = \nu_{1H} + |\nu_{15N}|$ (**fig. 2d**), where $\gamma_X$ is the magnetogyric ratio and $\nu_X$ is the corresponding Larmor frequency with X = $^1$H or $^{15}$N. Note, that $\gamma_{15N}$ is negative, so, the resulting frequency of the polarization oscillations is higher than $\nu_{1H}$.

Compared to SABRE-SHEATH $^{15}$N polarization, alt-SABRE-SHEATH provides significantly higher polarization levels (simulations, **fig. 2b vs. 2c,d**).

For experimental validation we used a SQUID based ultra-low field (ULF) NMR spectrometer,[24] allowing the simultaneous observation of all MR visible nuclei (in our case $^1$H and $^{15}$N). Although the chemical resolution is missing due to ULF detection at $B_0$ = 54 µT, the field homogeneity of the setup is sufficient to obtain J-resolved NMR spectra with a linewidth < 0.5 Hz.

The setup was proved to be very robust for long lasting SABRE experiments, which was demonstrated through ULF-SABRE-COSY lasting for more than 8 hours straight at room temperature and pressure with continuous pH$_2$ supply.[25]

A standard SABRE system consisting of [Ir] and $^{15}$N-acetonitrile dissolved in methanol, resulting in an active SABRE complex after activation with pH$_2$, was chosen for all following experiments (**fig. 2a**).

In a common SABRE-SHEATH experiment the system was flushed with pH$_2$ at a given field $B_{hyp}$ for a time period $t_{hyp}$, and then shuttled to the observation field (**fig. 1a**).[19] In our case, we simply increased $B_{hyp}$ after polarization to reach $B_0$ ~ 54 µT for ULF NMR observation. The maximum achievable polarization was at 1.2 µT (**SI, fig. S2**).

In the next step, the slow alt-SABRE-SHEATH oscillations near 1 µT were demonstrated for different $t_{high}$ and $B_{low}$ (**fig. 3**). The oscillations were visible for the $^{15}$N (**fig. 3a**) and $^1$H spectrum (**fig. 3b**). Figure **3c** shows the integrated real part, *ReI*, of the $^{15}$N spectrum, complemented with a sinusoidal fit of the polarization oscillation frequency $\nu$.

The lowest $\nu$ near $B_{LAC}$ ~ 1.2 µT was $\nu$ ~ 70 Hz (**SI, fig. S4c**) that increased to $\nu$ ~ 119 Hz at ~ 2.6 µT (**fig. 3c** and **SI, fig. S5c**) and $\nu$ ~ 185 Hz at ~ 3.9 µT (**SI, fig. S6c**). In the case of short-lived SABRE complexes, $\nu$ is equivalent to the rate of polarization transfer, meaning higher $\nu$ can achieve higher signal gain, despite the fact, that $B_{low}$ is not fulfilling the optimum $B_{LAC}$ condition.

Next, we kept both $t_{low}$ and $t_{high}$ shorter than $1/J_{HN}$ that resulted in an observation of a much faster oscillation with a frequency close to $\nu_{1H} + |\nu_{15N}|$ along the $t_{high}$ axes (**fig. 2d** and **fig. 4**). As discussed above, the frequency equals to $^1$H-$^{15}$N two-spin order zero quantum coherence at $B_0$. And, as expected, the oscillations were visible for the $^{15}$N (**fig. 4a**) and $^1$H signal (**fig. 4b**). Figure **4c** shows the *ReI* and fit of the $^{15}$N signal. The importance of multiple quantum coherences for SABRE experiments was shown in ref. [25].

This fast oscillation was observed for different parameter settings (**SI, fig. S7** and **S8**). Due to the long hyperpolarization build up times (> 10 s) and limited availability of the involved substances (acetonitrile, Ir-catalyst) only a very narrow parameter space of the performed simulations could be experimentally assessed. However, even these results confirm and reveal that although polarization is generated at low fields the evolution continues at high fields, where only refreshment of the Ir-complex was assumed. Here we show that the length of $t_{high}$ is also important for alt-SABRE-SHEATH hyperpolarization.

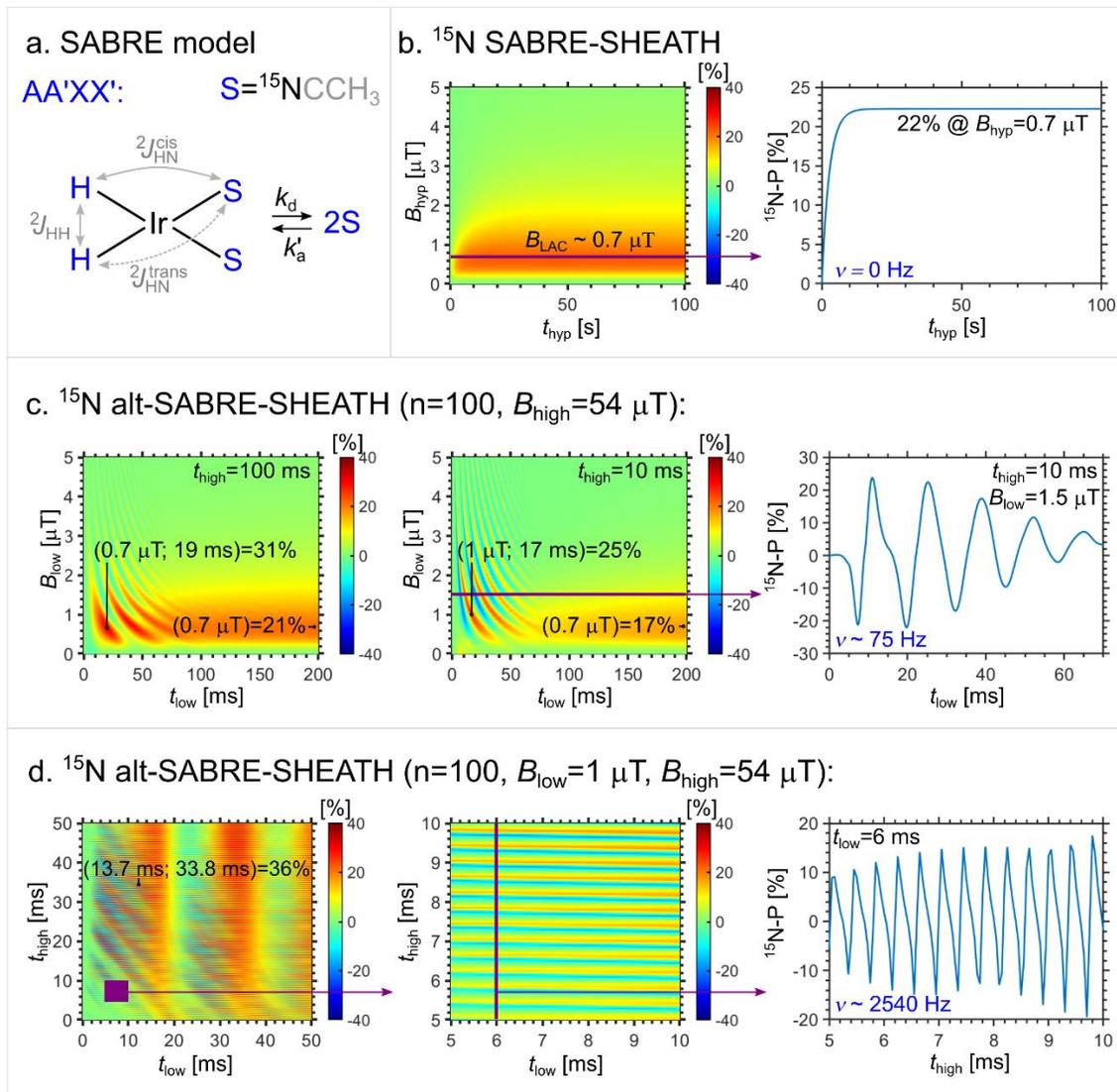

**Figure 2. Simulations of (alt-)SABRE-SHEATH experiments.** (a) AA'XX' system and SABRE exchange model.[21,22] (b) $^{15}$N SABRE-SHEATH (**fig. 1a**) polarization as a function of magnetic field $B_{hyp}$, hyperpolarization time $t_{hyp}$ and polarization build-up at $B_{hyp}$ = 1 µT. (c) $^{15}$N alt-SABRE-SHEATH (**fig. 1b**) polarization of S as a function of $B_{low}$ and $t_{low}$ for $t_{high}$ = 100 ms and 10 ms and an example of polarization oscillations. (d) $^{15}$N alt-SABRE-SHEATH polarization of S as a function of $t_{low}$ and $t_{high}$ for $B_{low}$ = 1 µT and $B_{high}$ = 54 µT. When both time periods are short the faster oscillations are visible (oscillation frequencies $\nu$ are given for the example kinetics – right plots). System parameters used for simulations were[21,22] $k'_a$ = 2 s$^{-1}$, $k_d$ = 20 s$^{-1}$, $t_{ramp}$ = 0, n=100; the complex consisted of two protons and two $^{15}$N with a high field $T_1$ of 1 s and 6 s, chemical shifts are -22 ppm and 0 ppm and $^2J_{HN}^{trans}$ = -24 Hz, $^2J_{HN}^{cis}$ = 0 and $^2J_{HH}$ = -8 Hz; after dissociation of $^{15}$N its high-field $T_1$ was 60 s and chemical shift 100 ppm. Two substrates S in the Ir-complex were considered; each consisted of one $^{15}$N. Note the variation of the frequency of polarization transfer in (c) when field deviates from $B_{LAC}$.

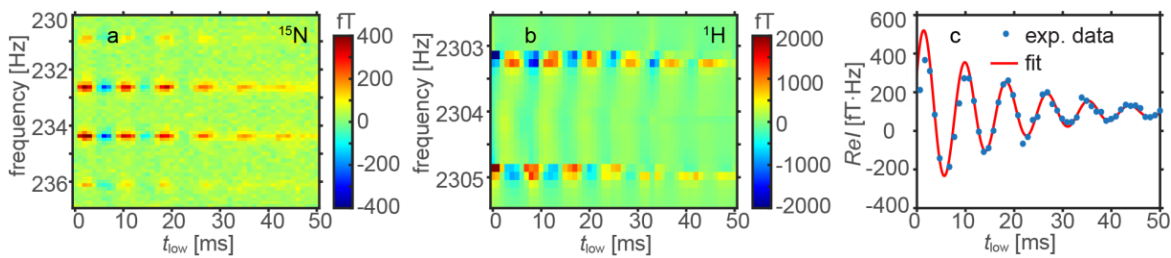

**Figure 3. Experimental alt-SABRE-SHEATH $^{15}$N and $^1$H spectra (a, b) and integral over $^{15}$N signal (c) as a function of $t_{low}$ at $B_{low}$ = 2.6 µT.** Note that both $^{15}$N and $^1$H signals of acetonitrile oscillate as a function of $t_{low}$ (these protons are neglected in the simulations). Other experimental parameters: $t_{hyp}$ = 30 s, $t_{high}$ = 10 ms, and $B_{high}$, $B_0$ = 54 µT. The kinetics (c) were fitted by $f = A + B\sin(2\pi\nu t_{low} + \phi)\,e^{-t_{low}/T_d}$ (fit, line) with fitting parameters: $\nu$ = 119.0 ± 1.1 Hz and $T_d$ = 17.0 ± 2.1 ms.

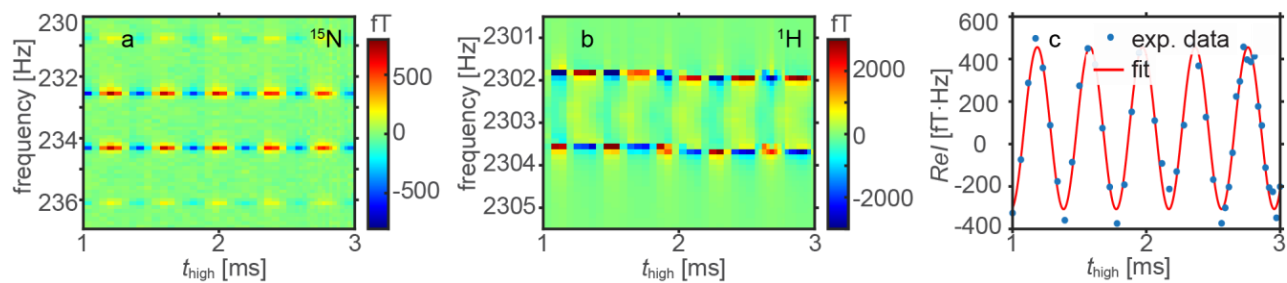

**Figure 4.** Experimental alt-SABRE-SHEATH $^{15}$N and $^1$H spectra (a, b) and integral over $^{15}$N signal (c) as a function of $t_{high}$ at $B_{high}$ = 54 µT. Note that both $^{15}$N and $^1$H signals of acetonitrile oscillate as a function of $t_{high}$ (these protons are neglected in the simulations). Other experimental parameters: $B_{low}$ = 2.6 µT, $t_{hyp}$ = 30 s, $t_{low}$ = 1.5 ms, and $B_0$ = 54 µT. The kinetics (c) were fitted by $f = A + B \sin(2\pi \nu t_{high} + \phi)$ (fit, line) with fitting parameters: $\nu = 2541 \pm 13$ Hz.

We were able to probe the short time dynamics, driven by second order quantum coherences of active SABRE complex and achieved a 30 % higher $^{15}$N polarization with alt-SABRE-SHEATH compared to common SABRE-SHEATH of equal $t_{hyp}$ (**SI 7**), in accordance with theory. This was accomplished using the sequence parameters depicted in **fig. 4**.

The analysis of alt-SABRE performance shows that maybe instead of non-adiabatic (fast in respect to LAC frequency splitting) variation of external magnetic fields it would be beneficial to use slower ramps, which will prevent such a fast and detrimental oscillation at $B_{high}$. Moreover, adiabatic passages through the LAC condition are sometimes beneficial for polarization transfer.[26,27]

The oscillations at $B_{low}$ with frequency $\nu$ (close to LAC condition) and at $B_{high}$ with frequency $\nu_{1H} + |\nu_{15N}|$ (weak $^1$H-$^{15}$N coupling) were uncovered experimentally and confirmed by simulations. A long pH$_2$ refreshing time ($t_{high}$) is beneficial for polarization transfer since the obtained signal oscillates around positive values (**fig. 3c** and **4c**). However it cannot be too long because then the relaxation starts playing a significant role.

For $t_{low}$ and $t_{high}$ shorter than $1/J_{HN}$ we showed that, a higher polarization than for common SABRE-SHEATH can be achieved. Since we only assessed a narrow parameter space further investigations are necessary to exploit this effect in order to gain maximum possible polarization.

The presented simulations and experiments were focused on $^{15}$N of acetonitrile. If and to what extend the polarization of other nuclei and other substrates can be boosted has to be investigated.

### Methods

**SABRE-SHEATH** (**fig. 1a, SI, fig. S1-S3**). $B_{hyp}$ was varied in the range of -0.5 to 3.5 µT to find $B_{LAC}$. pH$_2$ was flushed continuously. Close to 100 % enrichment pH$_2$ was prepared using in house built liquid helium pH$_2$ generator.[24]

**Alt-SABRE-SHEATH** (**fig. 3 and 4, SI, fig. S4-S8**). Alt-SABRE-SHEATH sequence (**fig. 1b**) has four experimental parameters, which were varied: $t_{low}$, $t_{high}$, $B_{low}$ and $B_{high}$. Ramp time $t_{ramp}$ was kept constant and equal to 10 ms. The alt-SABRE cycle was repeated $n$-times to reach a constant hyperpolarization time. Then a 90° RF-pulse flipped the $^{15}$N and $^1$H spins into the transversal plane for NMR signal observation. Since a broad band SQUID based detector was used, the $^{15}$N and $^1$H signals are read out simultaneously. Parahydrogen was bubbled through the reactor during the entire experiment. Since the variation of all parameters is very time consuming due to a long hyperpolarization period $t_{hyp}$ > 10 s only some projections of this 4-dimensional experimental space were measured (**fig. 3 and 4, SI, fig. S4-S8**). The alternating magnetic fields were created by a homemade current source that was controlled by an arbitrary waveform generator (model Rohde & Schwarz HM8150).

**Sample preparation.** The samples consisted of 40 µL acetonitrile-$^{15}$N (98% $^{15}$N, Merck, CAS: 14149-39-4) and 3 mg [IrImesCODCl][23] (in house synthesis) catalyst dissolved in 14 mL methanol (99%). The sample was conducted into a 2 mL reactor. Due to constant pH$_2$ bubbling the sample evaporated at a rate of ≈1 mL/h enabling a maximum measurement time of ≈ 10 h (signal stability, **SI, fig. S3**).

**ULF MRI setup.** The home-made ULF MRI setup consists of a broad band SQUID based detector, sitting inside a low noise fiberglass dewar.[28] A homogenous magnetic field is achieved by using a tetracoil with a sphere radius of 260 mm. A Helmholtz coil accomplishes RF transmission and spin rotation. The whole system sits inside a cubic shielding chamber consisting of two layers of µ-metal for shielding low frequency magnetic fields and one aluminum layer for shielding RF fields.

The volume of the hyperpolarization reactor is about 2 mL. It has an inlet for the pH$_2$ and in- and outlets to the adjacent reservoir. As soon as the sample level in the reactor decreases, new sample from the reservoir flows into the reaction chamber.[24]

**SABRE simulations.** SABRE experiments were simulated using MOIN spin library[29] and quantitative SABRE theory.[22] The effective association and dissociation exchange constants for the substrate were $k'_a$ = 2 s$^{-1}$, $k_d$ = 20 s$^{-1}$. The parameters of [IrH$_2$(S=$^{15}$N)$_2$] = AA'XX': high-field $T_1$ of $^1$H was 1 s for $^1$H and 6 s for $^{15}$N. Their respective chemical shifts were -22 ppm and 0 ppm. J-coupling constants were $J_{HN}^{2,trans}$ = -24 Hz, $J_{HN}^{2,cis}$ = 0 and $J_{HH}^2$ = -8 Hz. The parameters of free substrate ($^{15}$N) were: high field $T_1$ = 60 s and chemical shift 100 ppm.

**Author Contributions**

KB, ANP, NK: conceptualization, methodology, writing – original draft, preparation. KB, ANP, NK, MP: investigation. MP: [Ir] synthesis. KB, NK: experiments. ANP: simulations. KS, JBH: supervision. KS, JBH, ANP: funding acquisition. KS: provided the environment. All authors contributed to discussions, writing and helped interpreting the results and have given approval to the final version of the manuscript.

# Acknowledgements

We acknowledge funding from German Federal Ministry of Education and Research (BMBF) within the framework of the e:Med research and funding concept (01ZX1915C), DFG (HO-4602/2-2, HO-4602/3, GRK2154-2019, EXC2167, FOR5042, TRR287), Kiel University and the Faculty of Medicine. MOIN CC was founded by a grant from the European Regional Development Fund (ERDF) and the Zukunftsprogramm Wirtschaft of Schleswig-Holstein (Project no. 122-09-053).

**Keywords:** alt SABRE SHEATH • ULF NMR • pH$_2$ hyperpolarization • parahydrogen • Level anti-crossing • acetonitrile • SABRE

# Supporting materials for

# Coherent evolution of signal amplification by reversible exchange in two alternating fields (alt–SABRE)

Andrey N. Pravdivtsev*[a], Nicolas Kempf[b], Markus Plaumann[c], Johannes Bernarding[c], Klaus Scheffler, Jan-Bernd Hövener[a] and Kai Buckenmaier*[b]

[a] Dr. A. N. Pravdivtsev and Prof. Dr. J.-B. Hövener
Section Biomedical Imaging, Molecular Imaging North Competence Center (MOIN CC), Department of Radiology and Neuroradiology,
University Medical Center Kiel, Kiel University,
Am Botanischen Garten 14, 24114, Kiel, Germany
E-mail: andrey.pravdivtsev@rad.uni-kiel.de

[b] M. Sc. N. Kempf, Dr. K. Buckenmaier
High-Field Magnetic Resonance Center, Max Planck Institute for Biological Cybernetics,
Max-Planck-Ring 11, 72076, Tübingen, Germany
E-mail: kai.buckenmaier@tuebingen.mpg.de.

[c] Dr. M. Plaumann, Prof. Dr. Dr. Bernarding
Institute for Biometrics and Medical Informatics, Otto-von-Guericke University,
Building 02, Leipziger Str. 44, 39120, Magdeburg, Germany

Contents

# 1. Build-up of $^{15}$N-acetonitrile polarization in SABRE-SHEATH

The SABRE-SHEATH hyperpolarization build-up time $T_{buildup}$ was measured by changing the length of hyperpolarization $t_{hyp}$ followed by a FID readout at $B_0$ after a 90° $B_1$ pulse, which excited both $^1$H and $^{15}$N (fig. 1). The $^1$H and $^{15}$N signals were acquired simultaneously.

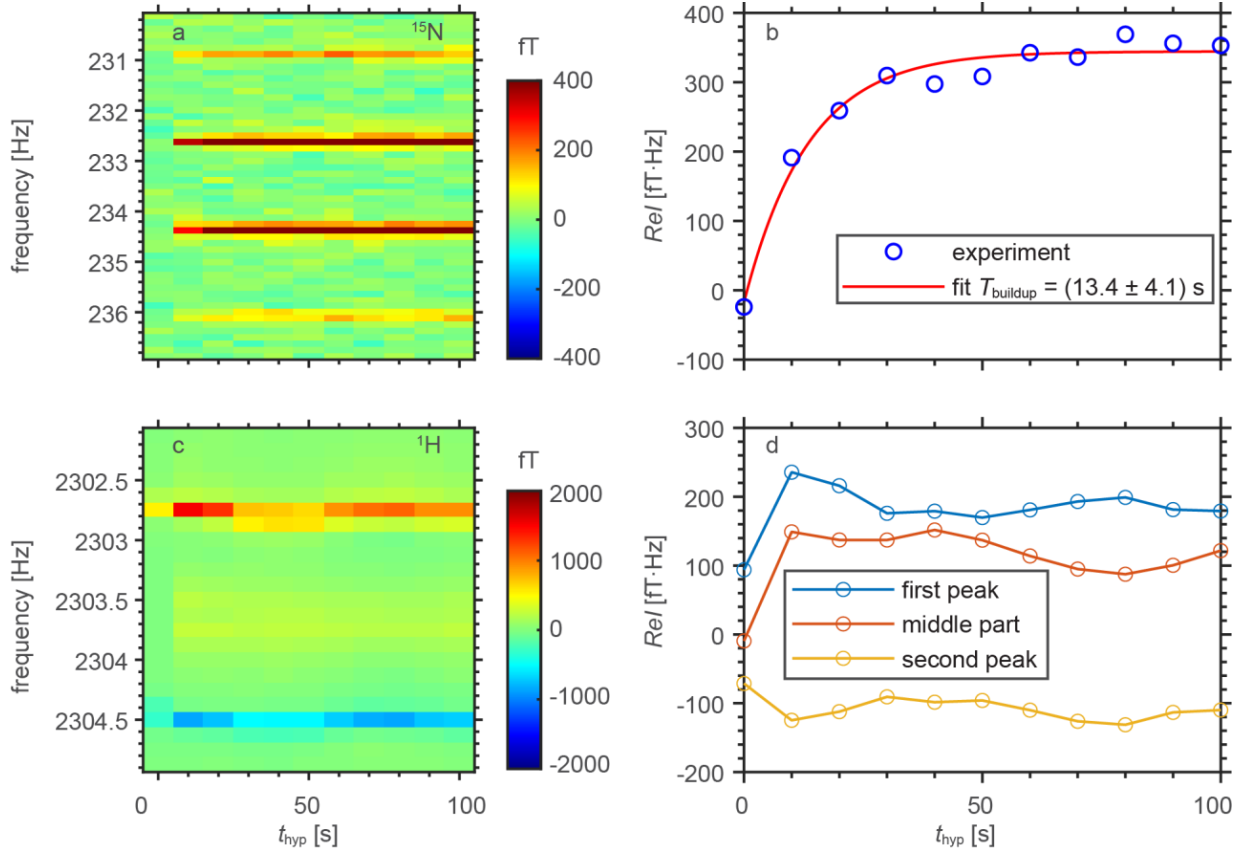

**Figure S1.** $^{15}$N (a and b) and $^1$H (c and d) signals of $^{15}$N-acetonitrile in SABRE-SHEATH as a function of $t_{hyp}$. In (b) the integral of the real part *ReI* (phased spectrum) is shown for the $^{15}$N signal together with an exponential fit. (d) shows the *ReI* for the two $^1$H peaks of $^{15}$N-acetonitrile and the middle part of the spectrum. The real part of the signal is integrated for three different areas, comprehensible in (c). For $^{15}$N-acetonitrile a doublet is expected. The first peak is defined as the peak at lower frequency, the second peak at higher frequency. The spectrum consists of an additional signal between the two main $^{15}$N-acetonitrile peaks, which is originating from free $H_2$ and Ir-complex.

**SABRE-SHEATH sequence parameters:** $B_{hyp} = B_{LAC} = 1.2$ µT; $t_{acq} = 8$ s; $B_{high} = B_0 = 54$ µT; $t_{ramp} = 10$ ms; 4 averages;

**Fitting function**:

$$ReI = P_{max}\left(1 - \exp(t_{hyp}/T_{buildup})\right) + offset \qquad (EQS1)$$

Where *ReI* is the integrated real part of the MR-signal.

**Fit parameters**:

$T_{buildup} = (13.4 \pm 4.1)$ s
$P_{max} = (361.9 \pm 48.0)$ fT·Hz
$offset = -(17.5 \pm 45.6)$ fT·Hz

## 2. Magnetic field dependency of $^{15}$N-acetonitrile polarization in SABRE-SHEATH

The $B_{low}$ dependency of the MR signal was measured via a simple SABRE-SHEATH hyperpolarization phase of fixed length followed by a FID readout (fig. 1).

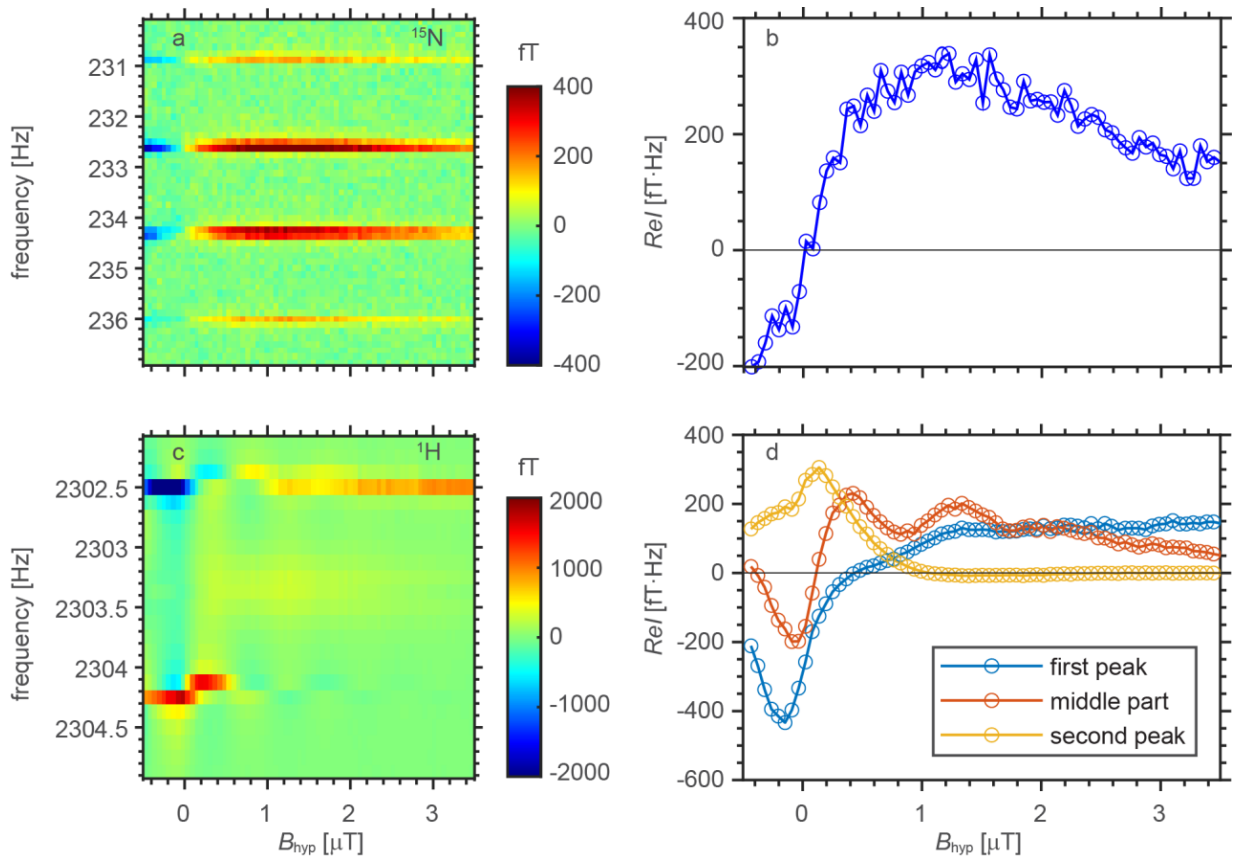

**Figure S2.** $^{15}$N (a and b) and $^1$H (c and d) signals of $^{15}$N-acetonitrile in SABRE-SHEATH as a function of $B_{hyp}$. In (b) the integral of the real part *ReI* is shown for the $^{15}$N signal. (d) shows the *ReI* for the two $^1$H peaks and the middle part. The maximum of the $^{15}$N signal was reached at $B_{hyp} \approx 1.2$ µT. This value is defined as $B_{LAC}$.

Note that the enhancement of the $^1$H signal behaves completely different compared to $^{15}$N. There is even hyperpolarization for $^1$H at $B_{hyp} = 0$.

**SABRE-SHEATH sequence parameters:** $t_{hyp}$ = 20 s; $t_{acq}$ = 8 s; $B_0$ = 54 µT; $t_{ramp}$ = 50 ms; 2 averages.

# 3. Stability of alt-SABRE-SHEATH during a one day experiment

During the alt-SABRE-SHEATH experiments presented in the main manuscript (fig. 3 and 4) the stability of the signal was measured via the integrated real part of the MR-signal, *ReI*, over the frequency range of the 4 $^{15}$N-acetonitrile peaks. We used a SABRE-SHEATH sequence with $t_{hyp}$ = 30 s. This enabled us to compare the magnitude of the $^{15}$N-signal in the SABRE-SHEATH and alt-SABRE-SHEATH experiments.

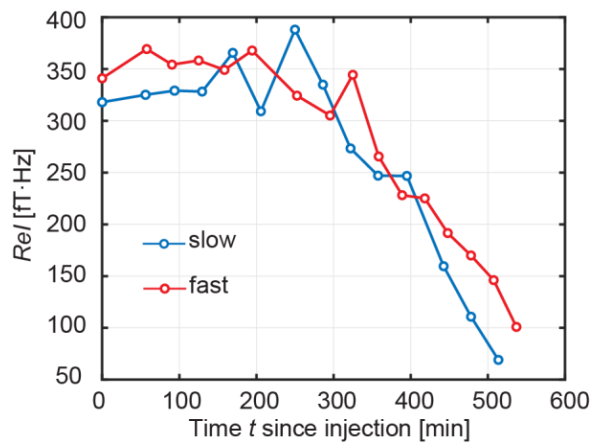

**Figure S3.** $^{15}$N-signal of $^{15}$N-acetonitrile for SABRE-SHEATH as a function of the total time of pH$_2$ flushing through the methanol solution of $^{15}$N-acetonitrile and Ir-catalyst. The SABRE-SHEATH stability measurements were interleaved with alt-SABRE experiments for slow (blue points, fig. 3) and fast (red points, fig. 4) oscillations presented in the main text.

**Sequence parameters:** $t_{hyp}$ = 30 s; $t_{acq}$ = 8 s; $B_{hyp}$ = $B_{LAC}$ = 1.2 µT; $B_0$ = 54 µT; $t_{ramp}$ = 10 ms; 4 averages.

## 4. Alt-SABRE-SHEATH at $B_{low}$ = 1.2 µT

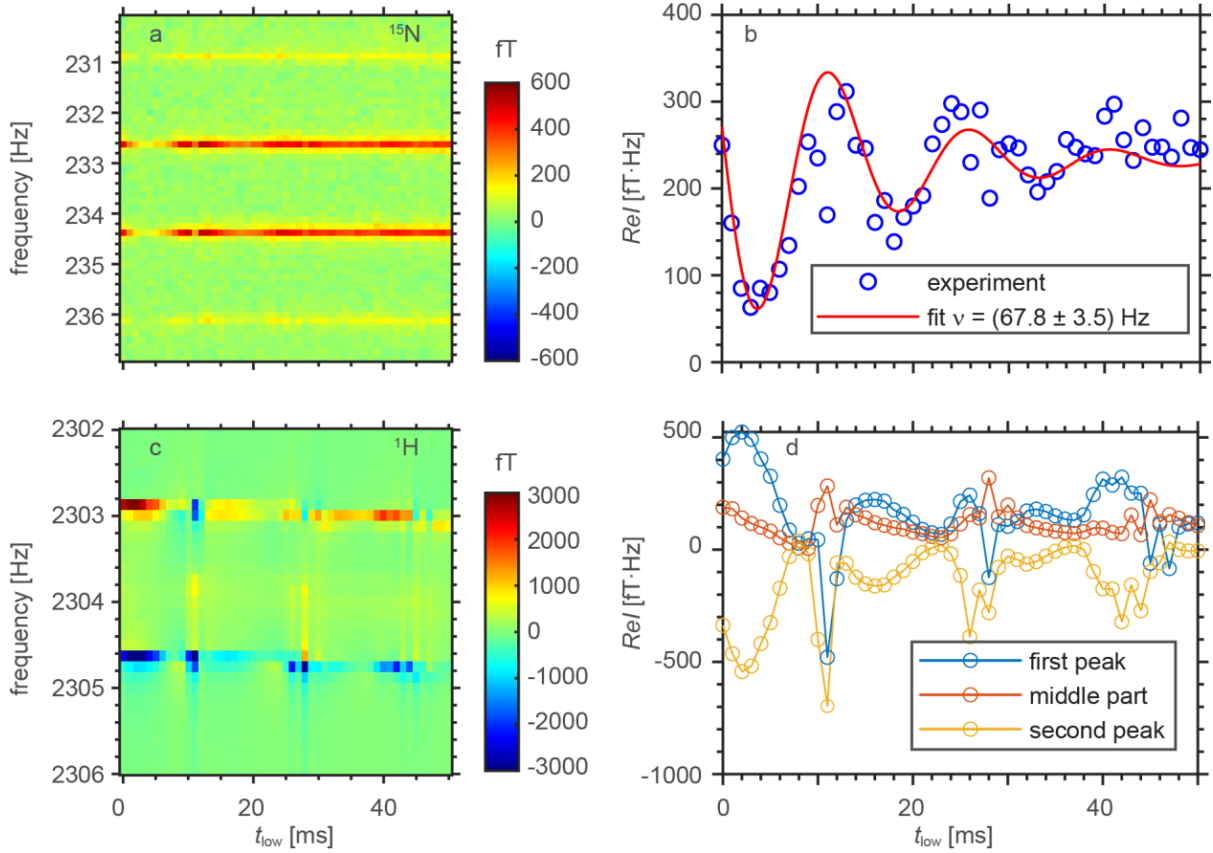

**Figure S4.** $^{15}$N (a and b) and $^1$H (c and d) signals of $^{15}$N-acetonitrile in alt-SABRE-SHEATH as a function of $t_{low}$ at $B_{low} = B_{LAC}$ = 1.2 µT. In (b) the integral of the real part *ReI* is shown for the $^{15}$N signal together with a sinusoidal fit. (d) shows the *ReI* for the two $^1$H peaks and the middle part. The fit for this dataset is slightly unmatched, since some additional effects occur (see spikes on the time dependences), when a maximum for the *ReI* of $^{15}$N is expected. The experiment was performed within ≈ 200 min. The signal stability during this period is given in fig. S3 in the range $0 < t < 200$ min.

**Sequence parameters:** $t_{hyp}$ = 30 s; $t_{acq}$ = 8 s; $t_{high}$ = 10 ms; $B_{low} = B_{LAC}$ = 1.2 µT; $B_{high} = B_0$ = 54 µT; $t_{ramp}$ = 10 ms; 4 averages.

Fitting function for *ReI* vs. $t_{low}$ (fig. S4b):

$$ReI = Amp \cdot \sin(\nu t_{low} + \varphi) \cdot \exp(-t_{low}/T_{decay} + \varphi) + offset \qquad (EQS2)$$

**Fit parameters:**

$\nu$ = (67.8 ± 3.5) Hz  
$T_{decay}$ = (13.9 ± 5.1) s  
$Amp$ = (226.9 ± 61.6) fT·Hz  
$offset$ = (232.8 ± 10.1) fT·Hz

# 5. Alt-SABRE-SHEATH at $B_{low}$ = 2.6 µT

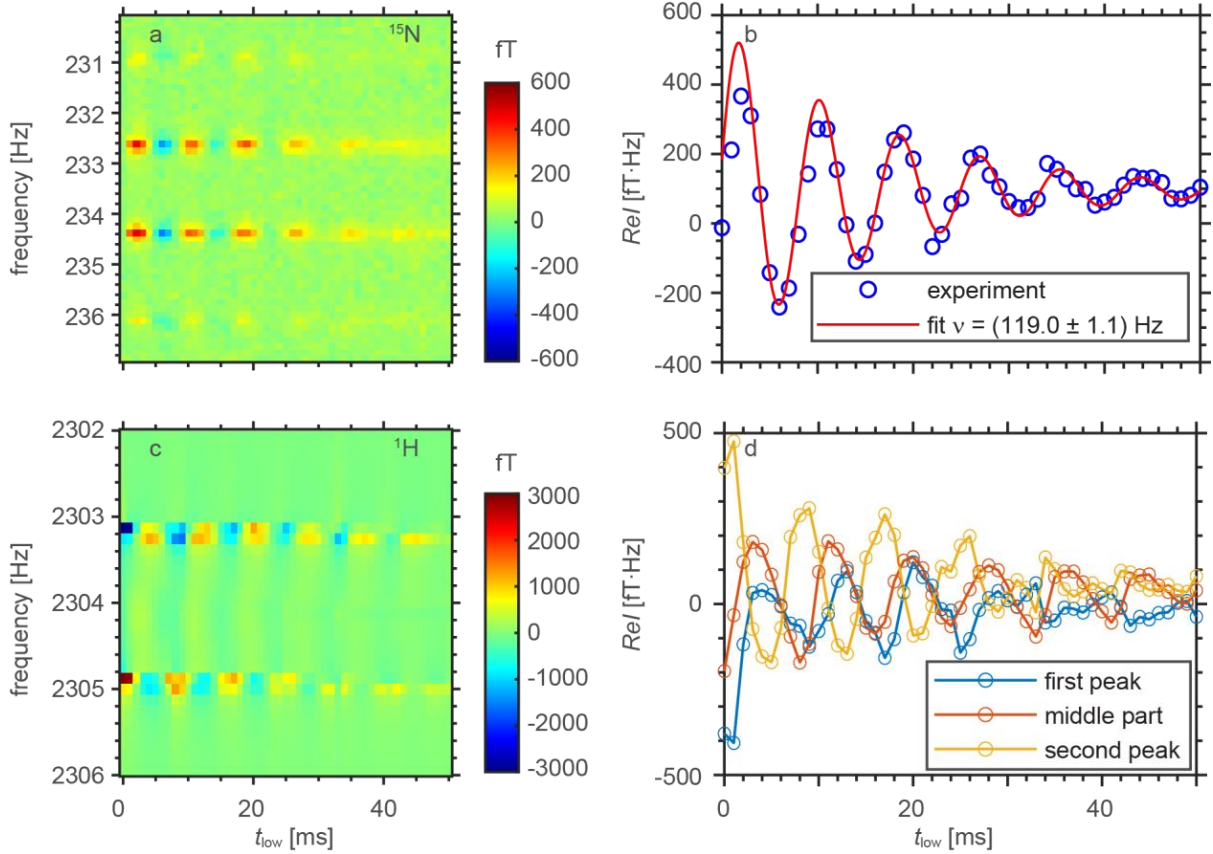

**Figure S5**. $^{15}$N (a and b) and $^1$H (c and d) signals of $^{15}$N-acetonitrile in alt-SABRE-SHEATH as a function of $t_{low}$ at $B_{low}$ = 2.6 µT. In (b) the integral of the real part *Re*I is shown for the $^{15}$N signal together with a sinusoidal fit. (d) shows the *Re*I for the two $^1$H peaks and the middle part. The experiment was performed within ≈ 200 min. The signal stability during this period is represented in fig. S3 in the range 200 < *t* < 400 min.

**Sequence parameters:** $t_{hyp}$ = 30 s; $t_{acq}$ = 8 s; $t_{high}$ = 10 ms; $B_{low}$ = 2.6 µT; $B_{high}$ = $B_0$ = 54 µT; $t_{ramp}$ = 10 ms; 4 averages.

Figure S5b was fitted with EQS2. The fit parameters are:

$\nu$ = (119.0 ± 1.1) Hz  
$T_{decay}$ = (17.0 ± 2.1) s  
*Amp* = (470.6 ± 39.6) fT·Hz  
*offset* = (97.1 ± 8.0) fT·Hz

## 6. Alt-SABRE-SHEATH at $B_{low}$ = 3.9 µT

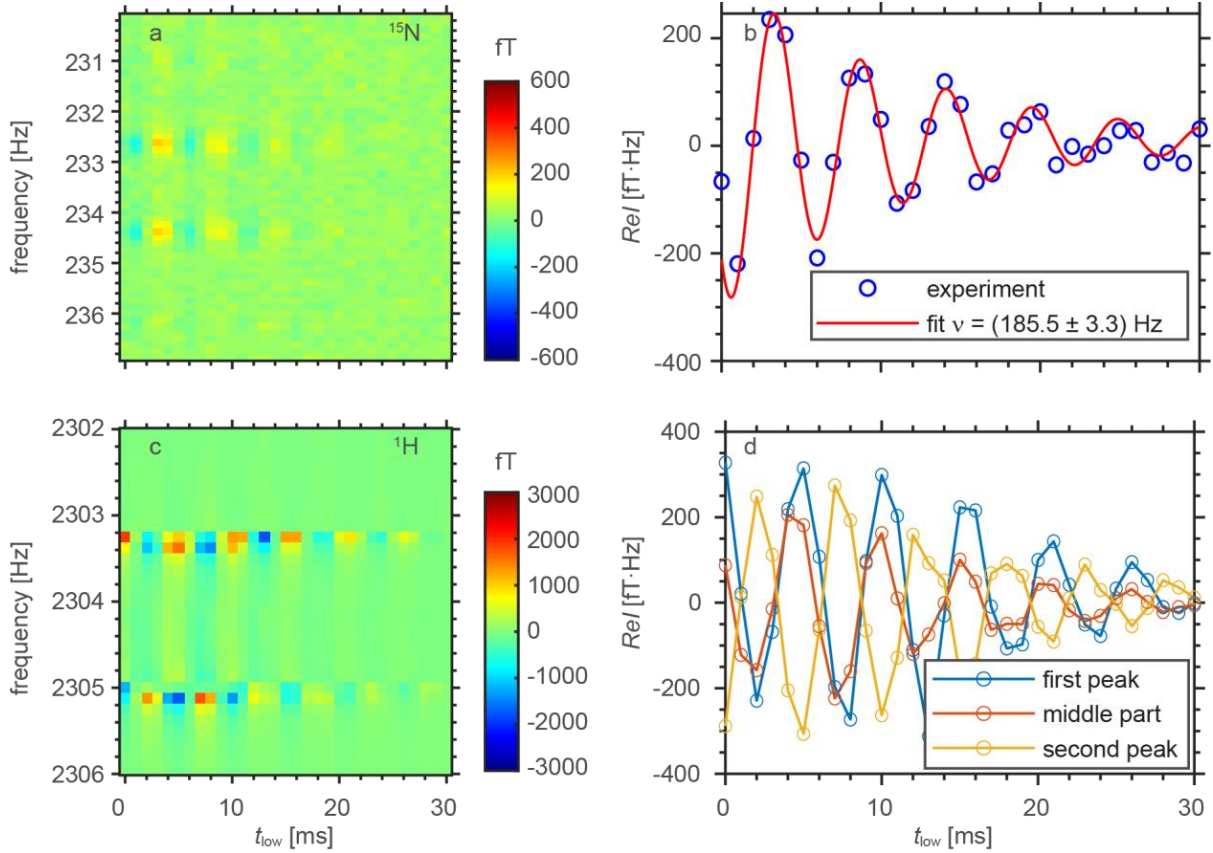

**Figure S6.** $^{15}$N (a and b) and $^{1}$H (c and d) signals of $^{15}$N-acetonitrile in alt-SABRE-SHEATH as a function of $t_{low}$ at $B_{low}$ = 3.9 µT. In (b) the integral of the real part *ReI* is shown for the $^{15}$N signal together with a sinusoidal fit. (d) shows the *ReI* for the two $^{1}$H peaks and the middle part. The experiment was performed within ≈ 120 min. The signal stability during this period is represented in fig. S3 in the range 400 < *t* < 520 min. As it can be seen, the signal intensity dropped significantly during this period. However, the increase in oscillation frequency compared to experiments with lower $B_{low}$ is clearly visible.

**Sequence parameters:** $t_{hyp}$ = 30 s; $t_{acq}$ = 8 s; $t_{high}$ = 10 ms; $B_{low}$ = 3.9 µT; $B_{high}$ = $B_0$ = 54 µT; $t_{ramp}$ = 10 ms; 4 averages.

Figure S6b was fitted with EQS2. The fit parameters are:

$\nu$ = (185.5 ± 3.3) Hz
$T_{decay}$ = (11.8 ± 2.8) s
*Amp* = (309.8 ± 47.2) fT·Hz
*Off* = (11.9 ± 11.1) fT·Hz

## 7. Alt-SABRE-SHEATH revealing fast oscillations

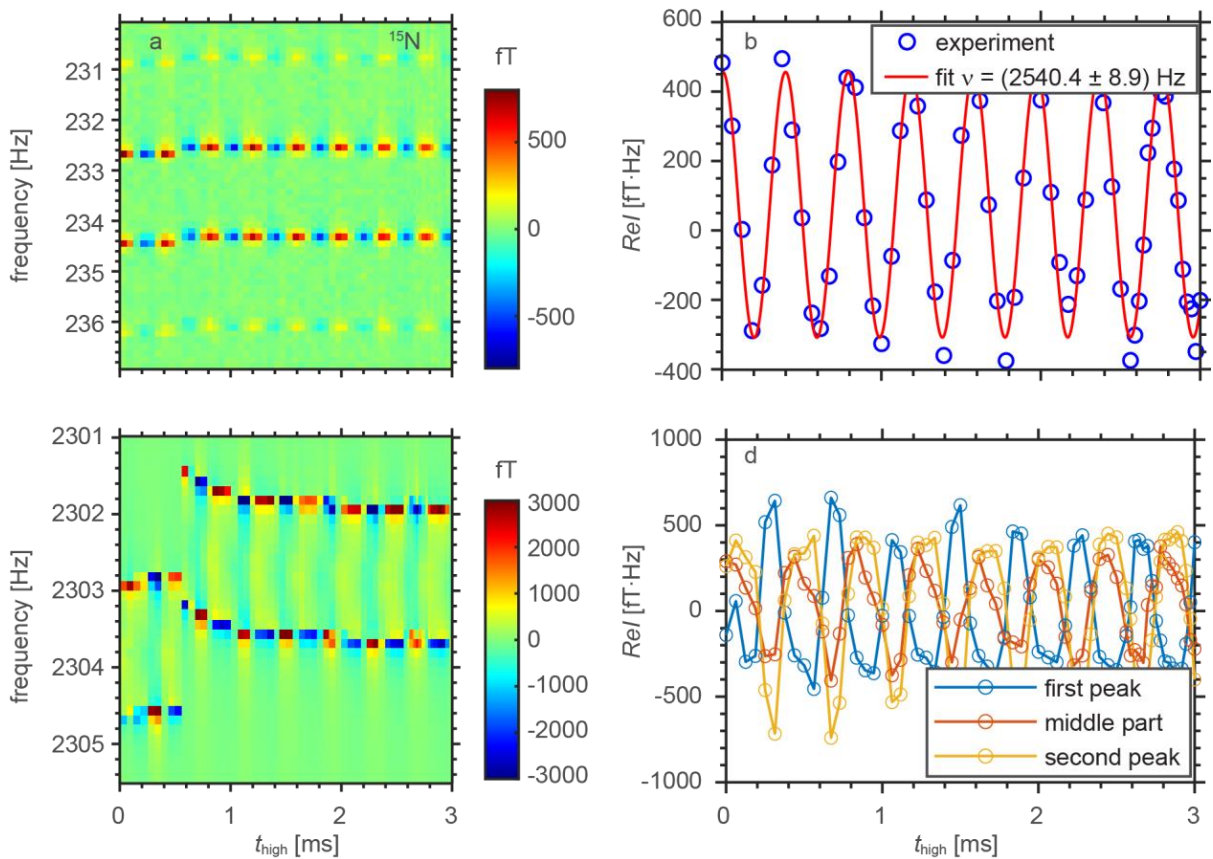

**Figure S7.** $^{15}$N (a and b) and $^1$H (c and d) signals of $^{15}$N-acetonitrile in alt-SABRE-SHEATH as a function of $t_{high}$ at $B_{high}$ = 54 µT. Much faster oscillations than at varying $B_{low}$ are present. In (b) the integral of the real part is shown for the $^{15}$N signal together with a sinusoidal fit. (d) shows the integral of the real part *ReI* for the two $^1$H peaks and the middle part. During the experiments stability measurements were acquired on a regularly basis (fig. S3). In order to do those measurements, the hardware had to be changed each time, which lead occasionally to $B_0$ jumps. For $^1$H it led to 10 times larger frequency variations than for $^{15}$N, because of the higher magnetogyric ratio of $^1$H compared to the one of $^{15}$N. Therefore in fig. S8c those jumps are notable the most. For simplicity, the main draft only shows the part without frequency jumps. The fit was performed over the whole scan, since the integral over the total frequency segment is only marginal influenced.

**Sequence parameters:** $t_{hyp}$ = 30 s; $t_{acq}$ = 8 s; $t_{low}$ = 1.5 ms; $B_{low}$ = $B_{LAC}$ = 1.2 µT; $B_{high}$ = $B_0$ = 54 µT; $t_{ramp}$ = 10 ms; 4 averages.

Fitting function for fig. S8b:

$$ReI = Amp \cdot \sin(\nu t_{high} + \varphi) + offset$$

**Fit parameters:**

$\nu$ = (2540.4 ± 8.9) s
*Amp* = (382.5 ± 18.8) fT·Hz
*offset* = (74.4 ± 13.5) fT·Hz

The maximum signal can be calculated via:

$P_{max,fO}$ = *Amp* + *Off* = (456.9 ± 23.1) fT·Hz

Note, that the average $P_{max,stab}$ for the stability measurements during the first 300 min was (352.0 ± 15.8) fT·Hz. This hints to a signal enhancement of ≈ 30 %.

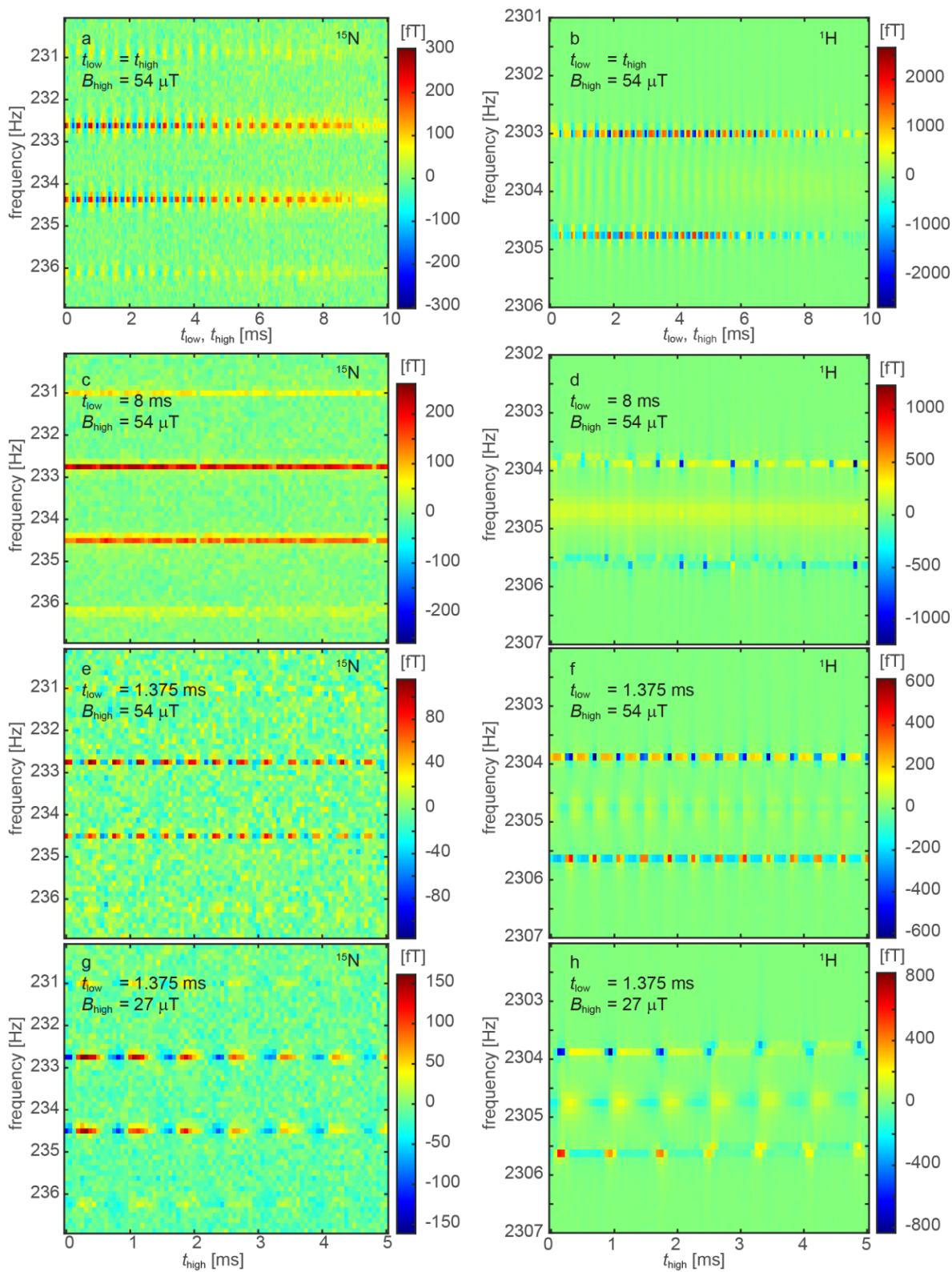

**Figure S8.** $^{15}$N (a, c, e and g) and $^{1}$H (b, d, f and h) alt-SABRE-SHEATH spectra as a function of $t_{high}$ at different conditions. In (a) and (b) $t_{low}$ and $t_{high}$ were varied simultaneously at $B_{high}$ = 54 µT. In (c) and (d) only $t_{high}$ was varied for a fixed $t_{low}$ = 8 ms. The same is shown in (e) and (f) for $t_{low}$ = 1.375 ms. Furthermore $B_{high}$ was lowered to 27.5 µT in (g) and (h). Because of the lower $B_{high}$, the oscillation frequency is lower, too. Experiments in fig. 8c – h were performed with a degenerated catalyst. Therefore the signal amplitude is lower in comparison to the other experiments.

**Sequence parameters for**

fig. S8a and b: $t_{hyp}$ = 10 s; $t_{acq}$ = 8 s; $t_{low}$ = $t_{high}$ = 0–10 ms; $B_{low}$ = $B_{LAC}$ = 1.2 µT; $B_{high}$ = $B_0$ = 54 µT; $t_{ramp}$ = 10 ms; 2 averages;

fig. S8c and d: $t_{hyp}$ = 10 s; $t_{acq}$ = 8 s; $t_{low}$ = 8 ms; $t_{high}$ = 0–5 ms; $B_{low}$ = $B_{LAC}$ = 1.2 µT; $B_{high}$ = $B_0$ = 54 µT; $t_{ramp}$ = 10 ms; 2 averages;

fig. S8e and f: $t_{hyp}$ = 10 s; $t_{acq}$ = 8 s; $t_{low}$ = 1.375 ms; $t_{high}$ = 0–5 ms; $B_{low}$ = $B_{LAC}$ = 1.2 µT; $B_{high}$ = $B_0$ = 54 µT; $t_{ramp}$ = 10 ms; 2 averages;

fig. S8g and h: $t_{hyp}$ = 10 s; $t_{acq}$ = 8 s; $t_{low}$ = 1.375 ms; $t_{high}$ = 0–5 ms; $B_{low}$ = 2.6 µT; $B_{high}$ = $B_0$ = 54 µT; $t_{ramp}$ = 10 ms; 2 averages;

Even so we have measured a qualitative agreement to simulations performed with a simplified AA'XX' 4 spin model, the experiments showed a more complex behavior as can be seen in fig. S1, S2, S4 and S8c and d. Figure S1 and S2 show unexpected fluctuations in the *Rel* of $^1$H as a function of $t_{hyp}$ and $B_{hyp}$. In the other figures there appear a drops or spikes in polarization. These spikes seem to have physical reasons, since the antisymmetric $^1$H signal changes to a symmetric one at the corresponding $t_{low}$, $t_{high}$ times and the absolute value of the $^1$H peak height is at a local maximum.

# 8. Energy level diagram for AA'XX'-type $^{15}$N-SABRE complex

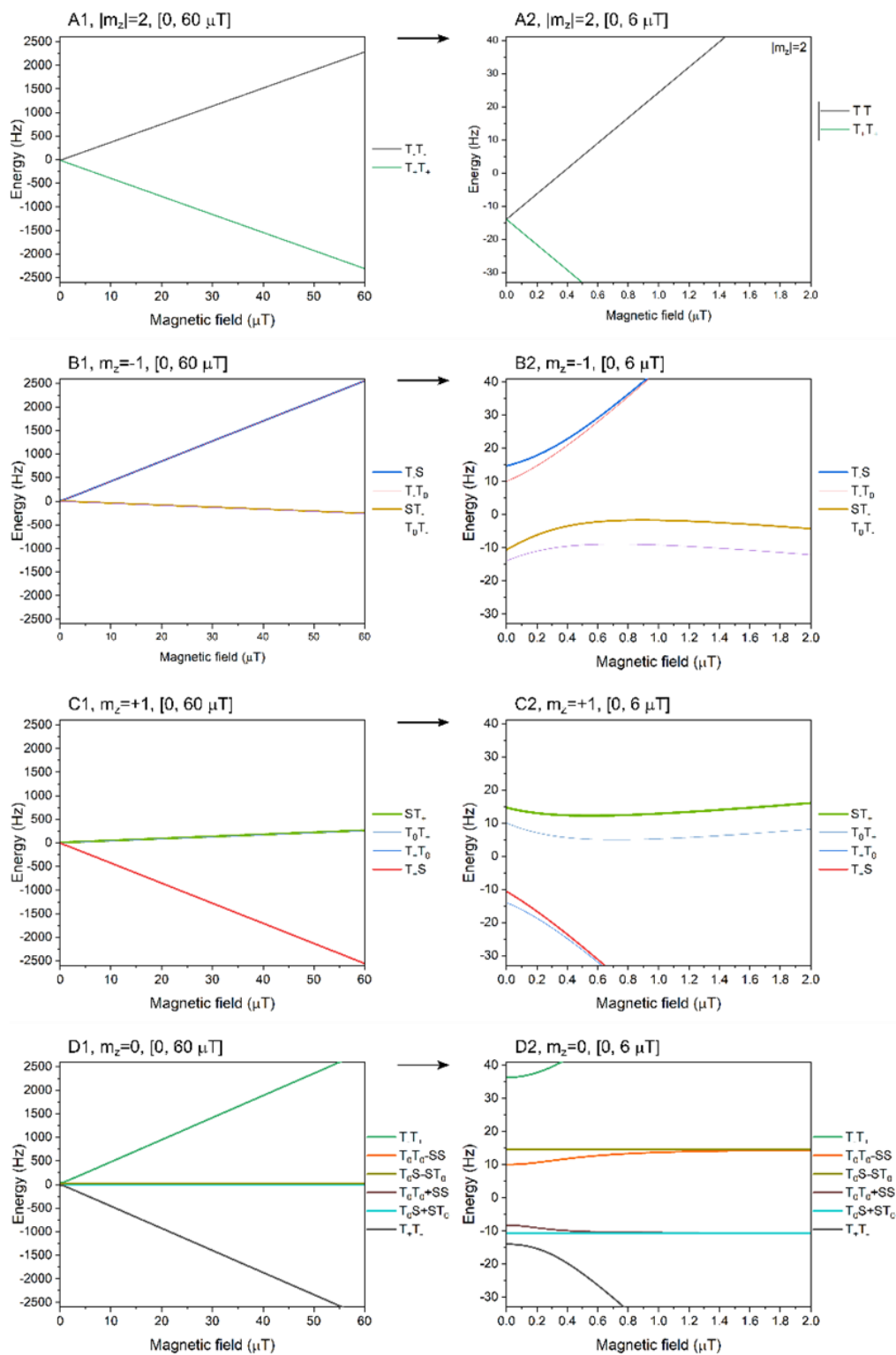

**Figure S9.** Energy diagrams of AA'XX' system of two $^1$H and two $^{15}$N in active SABRE complex for two field regions: [0, 60 µT] - left and [0, 2 µT] - right. The energy levels are grouped by the total projection of the spin state $m_z$: (A) $m_z$ = +2 or -2, (B) $m_z$ = -1, (C) $m_z$ = +1 and (D) $m_z$ = 0. On the low

field diagrams (right) various LACs are visible with energy level splitting ~ 100 Hz. At higher magnetic fields ~54 µT, the energy level splitting increases to 2500 Hz. The assignment of energy levels was done at high field of 10 T.